\documentclass[aps,prl,twocolumn,showpacs,floatfix,superscriptaddress,longbibliography]{revtex4-1}
\pdfoutput=1 % for arxiv, need to appear among the first 5 lines

\usepackage{amssymb,amsmath,amstext} %%   American Physical Society math etc extensions
\usepackage{graphicx}                %%   Include figure files  
\usepackage{epstopdf}                %%   help with eps -> pdf 
\usepackage{color}                   %%   change color
\usepackage{bm}                      %%   bold math                                    
\usepackage{appendix}                %%   formating appendicies
\usepackage[utf8]{inputenc}
\usepackage{bbold}
\usepackage{bbm}   
\usepackage{ulem}
\usepackage{latexsym}
\usepackage[colorlinks=true,allcolors=blue]{hyperref}
\usepackage{mathtools}

\usepackage[shrink=51,letterspace=800]{microtype}

\usepackage[euler]{textgreek}
\usepackage{upgreek}

\usepackage[dvipsnames]{xcolor}
\definecolor{dgreen} {RGB}{0,100,0}
\renewcommand{\vec}[1]{\mathbf{#1}}
\newcommand{\h}[1]{\hat{#1}}

\allowdisplaybreaks

\begin{document} 

\title{Borromean supercounterfluidity}
 \author{Emil Blomquist}
 \affiliation{Department of Physics, KTH Royal Institute of Technology, Stockholm SE-10691, Sweden}

\author{Andrzej Syrwid} 
 \affiliation{Department of Physics, KTH Royal Institute of Technology, Stockholm SE-10691, Sweden}

 \author{Egor Babaev}
 \affiliation{Department of Physics, KTH Royal Institute of Technology, Stockholm SE-10691, Sweden}

\begin{abstract}
We demonstrate microscopically the existence of a new superfluid state of matter in a three-component Bose mixture trapped in an optical lattice. The superfluid transport involving co-flow of all three components is arrested in that state, while counterflows between any pair of components are dissipationless. The presence of three components allows for three different types of counterflows with only two independent superfluid degrees of freedom.
\end{abstract}

\date{\today}

%%%%%%%%%%%%%%%%%%%%%%%%%%%%%%%%%%%%%%%%%%%%%%%%%%%%%%%%%%%%%%%%%%%%%%%%%%%%%%%%%%%%%%%

\maketitle
The advent of optical lattices~\cite{jaksch1998cold, jaksch2005cold, greiner2002quantum, bloch2005ultracold, bloch2008many, lewenstein2007ultracold} allowed for highly controllable access to strongly-correlated quantum many-body systems and opened up a way to realize various phases of matter.
One of the theoretical predictions was that bosons in optical lattices could host a new  type of transport phenomenon called supercounterfluidity~\cite{Kuklov2003counterflow, kuklov2004commensurate, kuklov2004superfluid, kuklov2006deconfined, capogrosso2008monte, dahl2008preemptive, dahl2008unusual, Dahl2008hidden,  capogrosso2011superfluidity,sellin2018superfluid}.
That is, having two bosonic fields $\psi_{1,2}$, in an ordinary case, one finds a superfluid mixture when $\langle \psi_{1,2}\rangle$ reveals a (quasi) long-range order.
This  state is predicted to appear ~\cite{Kuklov2003counterflow, kuklov2004commensurate, kuklov2004superfluid, kuklov2006deconfined, capogrosso2008monte, dahl2008preemptive, dahl2008unusual, Dahl2008hidden,  capogrosso2011superfluidity,sellin2018superfluid} when the averages of individual fields vanish, $\langle\psi_{1,2}\rangle=0$, but there is a (quasi) long-range order in the composite field $\langle\psi_1\psi_2^*\rangle \ne 0$.
Therefore, in a supercounterfluid phase, individual bosonic species do not exhibit superfluidity, but the transport of particle-hole composites is dissipationless.
A similar type of order was predicted in superconducting systems arising from a different microscopic origin~\cite{babaev2002phase,babaev2004superconductor,smiseth2005field,herland2010phase,kuklov2008deconfined,agterberg2008dislocations,berg2009charge,weston2021composite}. 
For a general discussion, see~\cite{svistunov2015superfluid}. Recently an experimental observation of a discrete-symmetry counterpart of this type of order was reported in a three-component superconductor~\cite{grinenko2021bosonic}.
In the two-component case, the order parameter for the supercounterfluid is partially similar to a condensate of bound particle-hole pairs between two different components. However, the three-component situation is more subtle. 
In this Letter, we demonstrate microscopically the existence of a new ``super" state in a three-component Bose mixture.

At the superfluid hydrodynamic level, $N$-component isotropic superfluid systems can in general be described by the free-energy density $ f=\frac{1}{2} \sum_{\alpha,\beta}\rho_{\alpha\beta}\vec v_\alpha \cdot \vec v_\beta $.
Here $\vec v_\alpha = \nabla \theta_\alpha/m_\alpha$ ($\hbar=1$) represents the superfluid velocity of component $\alpha$, where $\theta_\alpha\in[0,2\pi)$ and $m_\alpha$ denote superfluid phase and particle mass of the $\alpha$\textsuperscript{th} component, respectively. 
In the presence of inter-component interactions, the superfluid stiffness tensor $\rho$ will in addition to the diagonal superfluid density elements, $\rho_{\alpha\alpha}$,  also contain off-diagonal elements, $\rho_{\alpha\beta}$ ($\alpha\neq\beta$), coupling the different components~\cite{andreev1976three}. This in turns has principal consequences for the corresponding superflows $ \vec j_\alpha=\partial f/\partial \vec v_\alpha = \rho_{\alpha\alpha} \vec v_\alpha + \sum_{\beta\neq \alpha} \rho_{\alpha\beta} \vec v_\beta $. Namely, the superflow of one component can be induced by the flow of a different component.
This fundamental phenomenon is referred to as Andreev--Bashkin drag~\cite{andreev1976three}.
  The Andreev--Bashkin effect was studied  in various regimes in optical lattices for binary~\cite{Kuklov2003counterflow, kuklov2004commensurate, kuklov2004commensurate, capogrosso2008monte, linder2009calculation, hofer2012superfluid,contessi2021collisionless, sellin2018superfluid}, and trinary \cite{Hartman2018} mixtures.
Note that in certain asymmetrical optical lattices there are additional terms responsible for transverse entrainment~\cite{vecdrag}.
However, in this Letter, we will restrict ourselves to square lattices where only the Andreev--Bashkin drag effect is present.

In the simplest case of two components with identical masses, the free-energy density can be cast into the form associated with co- and counter-flows $ f = \uprho_{2}(\nabla \theta_1 +\nabla \theta_2)^2/4 + \uprho_{0}(\nabla \theta_1 - \nabla \theta_2)^2/4$.
Here $ \uprho_\xi =\rho_\text{k}+(\xi-1) \rho_\text{d}\ge 0 $
where $\rho_\text{k}>0$ describe the prefactor of the standard gradient term, and $\rho_\text{d}$---either positive or negative---denotes the drag strength.
When the drag is sufficiently strong and negative,  the cheapest topological excitations become co-circulating composite vortices, i.e., vortices where both phases $ \theta_{1,2} $ wind by $ 2\pi $ around the core~\cite{svistunov2015superfluid}. Thermal or quantum fluctuations can then lead to the proliferation of these composite vortices---but not elementary ones---resulting in a phase transition to a supercounterfluid (for a detailed discussion of the principle, see, e.g., Chapter 6 in~\cite{svistunov2015superfluid}).
The composite vortices do not induce gradients in the phase difference and thus do not disorder the phase difference part of the free energy.
The free-energy density of the resulting state therefore only involves the phase stiffness corresponding to the phase difference, i.e., $f = \uprho_0 (\nabla \theta_1 - \nabla \theta_2)^2 / 4$.
This term can be interpreted as the kinetic free-energy contribution related to the composite particle-hole order parameter $\psi_1\psi_2^*$.
Consequently, only counter-flow dissipationless transport can take place.

In a system with more than two components, states may arise with no direct counterpart among two-component superfluids. 
Let us therefore consider a two-dimensional $N$-component symmetric quantum system, i.e., components with identical masses and densities, and equal Andreev--Bashkin drag strength  $\rho_\text{d}$ between each pair. 
We start with a phase-only approximation assuming identical and homogeneous densities of unit mass particles ($m_\alpha = 1$) in each superfluid component. 
The corresponding free-energy density reads
\begin{equation} \label{action}
\begin{split}
    f &
    =
    \frac{\rho_\text{k}}{2}\sum_\alpha (\nabla\theta_\alpha)^2
    +
    \frac{\rho_\text{d}}{2}
    \sum_{\alpha,\beta\neq\alpha}
    \nabla \theta_\alpha \cdot \nabla \theta_\beta
    \\
    &=  
    \frac{\uprho_N}{2N}\bigg(\sum_\alpha \nabla \theta_\alpha\bigg)^2
    +
    \frac{\uprho_0}{4N}\sum_{\alpha,\beta}(\nabla \theta_\alpha - \nabla \theta_\beta)^2
    \, ,
\end{split}
\end{equation}
where again $ \uprho_\xi =\rho_\text{k}+(\xi-1) \rho_\text{d}\ge 0$.
Now, when the drag is strong and negative, i.e., when $\uprho_N \ll \uprho_0$, the cheapest topological excitations are composite vortices where all $N$ phases wind by $2\pi$ with the same orientation. Consequently, the proliferation of the three-component topological defects leads to a state where the sum of three phases is disordered. However, these composite vortices are unable to disorder phase differences and the system retains
$N-1$ superfluid modes. Equation~(\ref{action}) therefore reduces to $ f \propto \sum_{\alpha,\beta}(\nabla \theta_\alpha - \nabla \theta_\beta)^2 $ and the corresponding phase is characterized by zero net superflow, i.e., $ \sum_\alpha \vec j_\alpha \propto \sum_\alpha  \partial f/\partial(\nabla \theta_\alpha) = \mathbf 0$.

\begin{figure}[t]
    \includegraphics[width=0.5\linewidth]{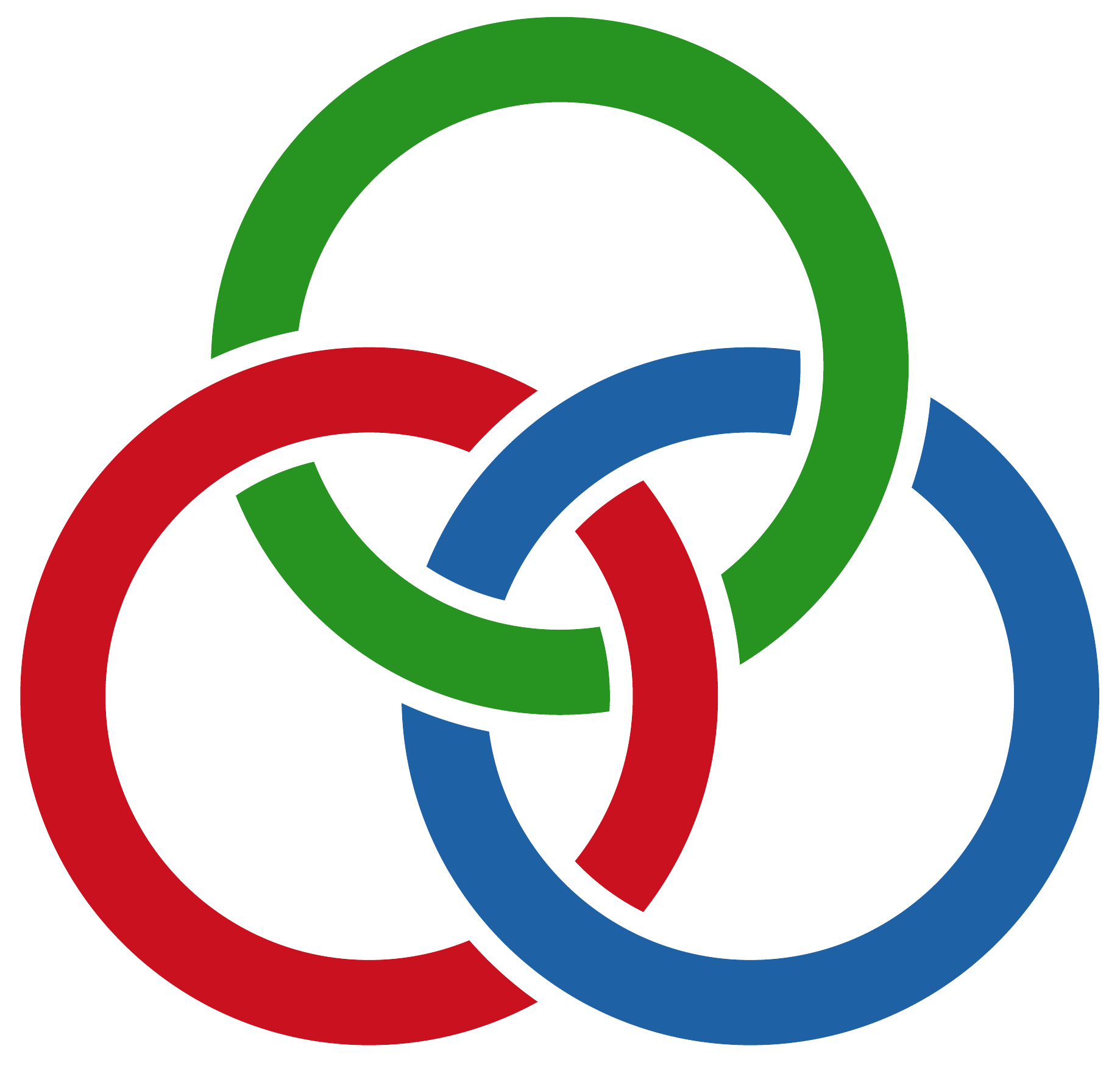}
    \caption{
         The Borromean rings. If a single ring is removed, the two remaining ones will become unlinked.
    }
    \label{fig:Borromean-rings}
\end{figure}

Consequently, for $ N > 2 $, the transport properties of the new phase can be understood as a counterflow of two components where the presence of the third symmetric component allows for fluctuations in the type of counterpropagating companions.
One would anticipate that this should be reflected in the world lines of the microscopic path-integral formulation. That is, in different regions of the system, one should find different types of particle-hole paired world lines.
Moreover, there is no superfluid co-flow of bound $ N $ particle states, while counter-propagation of any two different components is dissipationless.
Specifically, for $N=3$ where $\alpha\in\{\text{r},\text{g},\text{b}\}$, there are three types of counterflows for which there are only two independent degrees of freedom. 
That implies that we can have a superfluid co-flow of two components as long as their combined flow is counteracted by the flow of the third component, e.g., $ \vec{j}_\text{r} = \vec{j}_\text{g} = \vec{j} $ and $ \vec{j}_\text{b} = -2\vec{j} $.
Here we can draw a distant analogy to the Borromean rings where three rings are confined while each pair of rings is deconfined, see Fig.~\ref{fig:Borromean-rings}. Hence we coin this   phenomenon  Borromean supercounterfluidity.

Below we microscopically demonstrate that such superfluid state exists in a three-component ($ N = 3 $) Bose--Hubbard model~\cite{gersch1963quantum}
\begin{equation} \label{eq:BH3}
\begin{split}
    \h H
    =
    &-t
    \sum_\alpha\sum_{ \langle ij \rangle}
    \h b^\dagger_{i\alpha} \h b_{j\alpha}
    +
    \frac U 2
    \sum_\alpha \sum_i
    \h n_{i\alpha} \big( \h n_{i\alpha} - 1 \big)
    \\
    &+\frac{U'}{2}\sum_{\alpha, \beta \neq \alpha}\sum_i   \h n_{i\alpha} \h n_{i\beta}
    \,.
\end{split}
\end{equation}  
Here $ \h b_{i\alpha} $ ($ \h b^\dagger_{i\alpha} $) is the bosonic annihilation (creation) operator of component $ \alpha $ at site $ i $, and $ \h n_{i\alpha} = \h b^\dagger_{i\alpha} \h b_{i\alpha} $ is the corresponding particle number operator.
Greek subscripts label the component type, i.e., $ \text{r} $ (red), $ \text{g} $ (green), and $ \text{b} $ (blue).
The parameter $ t $ represents the hopping amplitude, while $ U $ and $ U' $, respectively, are the intra-component and inter-component on-site interaction strengths. 
We will consider a $ L \times L $ square lattice with unit lattice constant and periodic boundary conditions. 
We further analyze the two separate cases where either the individual particle-number densities are fixed, i.e., $ n_\alpha \coloneqq \langle \sum_i \h n_{i\alpha} \rangle / L^2 = 1 / 3 $, or the total particle-number density is conserved, i.e., $ \sum_\alpha n_\alpha = 1 $, while allowing for fluctuations in $n_\alpha$.

We numerically investigate the system by utilizing worm-algorithm Monte Carlo~\cite{prokof1998worm,capogrosso2007phase,Blomquist1590738,SellinThesis,lingua2018multiworm}---a quantum Monte-Carlo method which samples path-integral configurations of the partition function in real space and imaginary time. To extract the numerical values of $\rho_\text{k}$ and $\rho_\text{d}$ appearing in the free-energy density, Eq.~(\ref{action}), 
we generalize Pollock and Ceperley's formula~\cite{pollock1987path, sellin2018superfluid,vecdrag}: $ \rho_\text{k} = T \langle \vec w_\alpha^2 \rangle / 2 $, and $ \rho_\text{d} = T \langle \vec w_\alpha \cdot \vec w_\beta \rangle / 2 $ where $ T $ is the temperature ($k_\text{B}=1$) and $ \alpha \neq \beta $. 
The winding numbers $ \vec w_\alpha $ encode the net number of times, and in which direction, $\alpha$-type particles cross the periodic boundaries. The notation $ \langle \cdot \rangle $ refers to the standard statistical Monte Carlo average.

\begin{figure}[t]
    \includegraphics[width=1\linewidth]{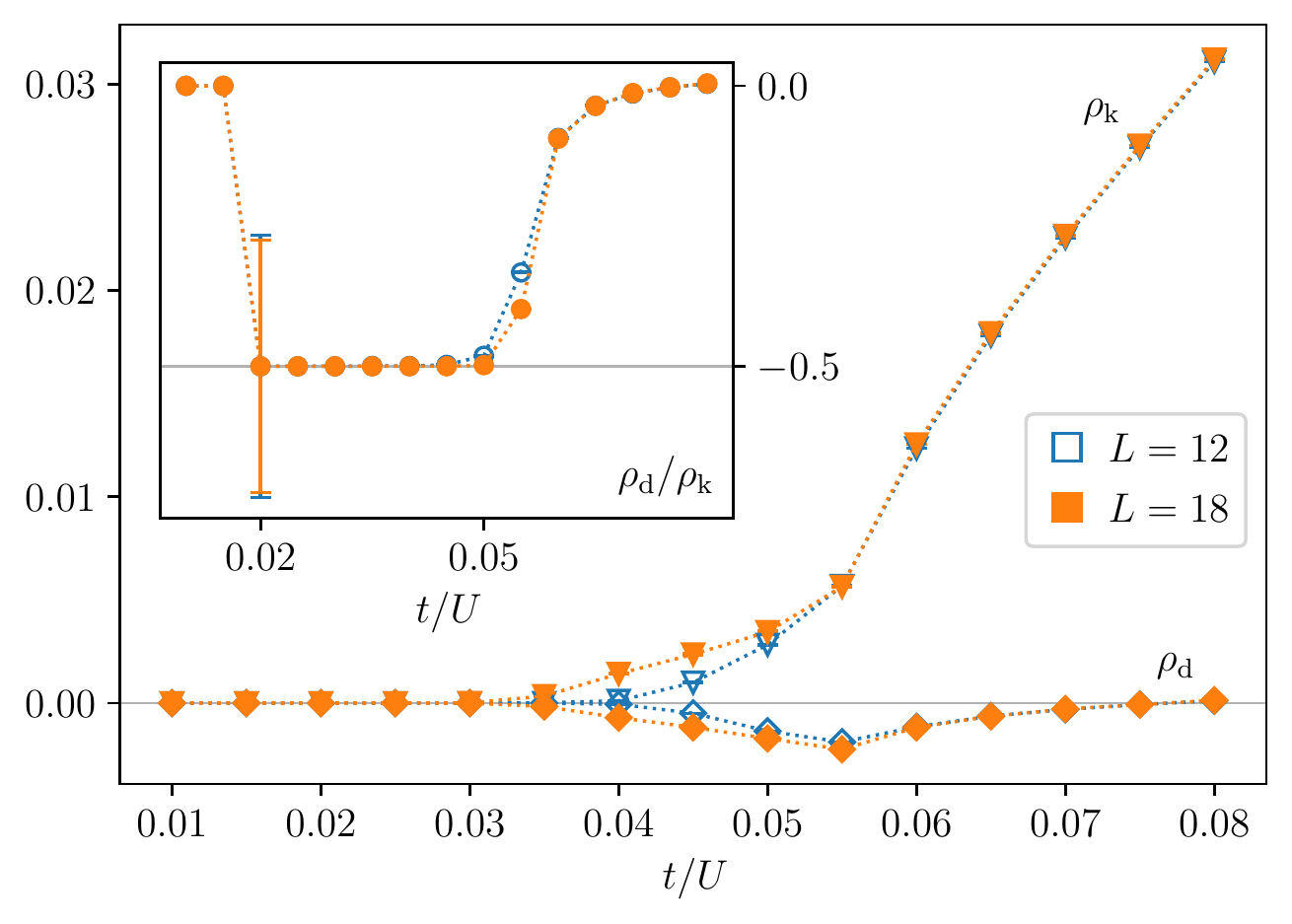}
    \caption{
    The superfluid density $ \rho_\mathrm{k} $ (triangles), the Andreev--Bashkin drag-coefficient $ \rho_\mathrm{d} $ (diamonds), and the ratio $ \rho_\mathrm{d} / \rho_\mathrm{k} $ (inset) as a function of $ t/U $.
    For $ t/U \in [0.02, 0.05] $ the ratio $ \rho_\mathrm{d} / \rho_\mathrm{k} $ saturates to its lower bound, $ -1/2 $, which indicates the presence of the Borromean supercounterfluid phase.
    The computations were performed using $U=1 $, $ U' = 0.9 $, $ T = t/L $, $ L = 12, 18 $, and a fixed particle number density $ n_\alpha = 1/3 $.
    }
    \label{fig:order_params}
\end{figure}

The calculated coefficients $ \rho_\text{k} $, $ \rho_\text{d} $, and their ratios $ \rho_\text{d} / \rho_\text{k} $ are presented in Fig.~\ref{fig:order_params} as functions of $ t/U $ for the interactions strengths $ U = 1 $ and $ U' = 0.9 $.
For small $ t/U $ we observe $ \rho_\text{k} = \rho_\text{d} = 0 $, which indicates a Mott insulating phase. However, at  $ t/U \gtrsim 0.02 $ the system enters the Borromean supercounterfluid phase, where $\rho_\text{d}/\rho_\text{k}$ rapidly saturates at the value $-1/2$ for which the coefficient $\uprho_{N=3}$ in Eq.~(\ref{action}) vanishes. This result demonstrates that the corresponding effective free-energy density is given by gradients of phase differences between all the three components. When further increasing  $t/U \gtrsim 0.05$, the system undergoes a second transition to the three-component superfluid phase  where the $ U(1)\times U(1) \times U(1) $ symmetry is broken. Deep in this regime $ \rho_\text{d}/\rho_\text{k}$ becomes very small.

\begin{figure}[t]
    \includegraphics[width=1\linewidth]{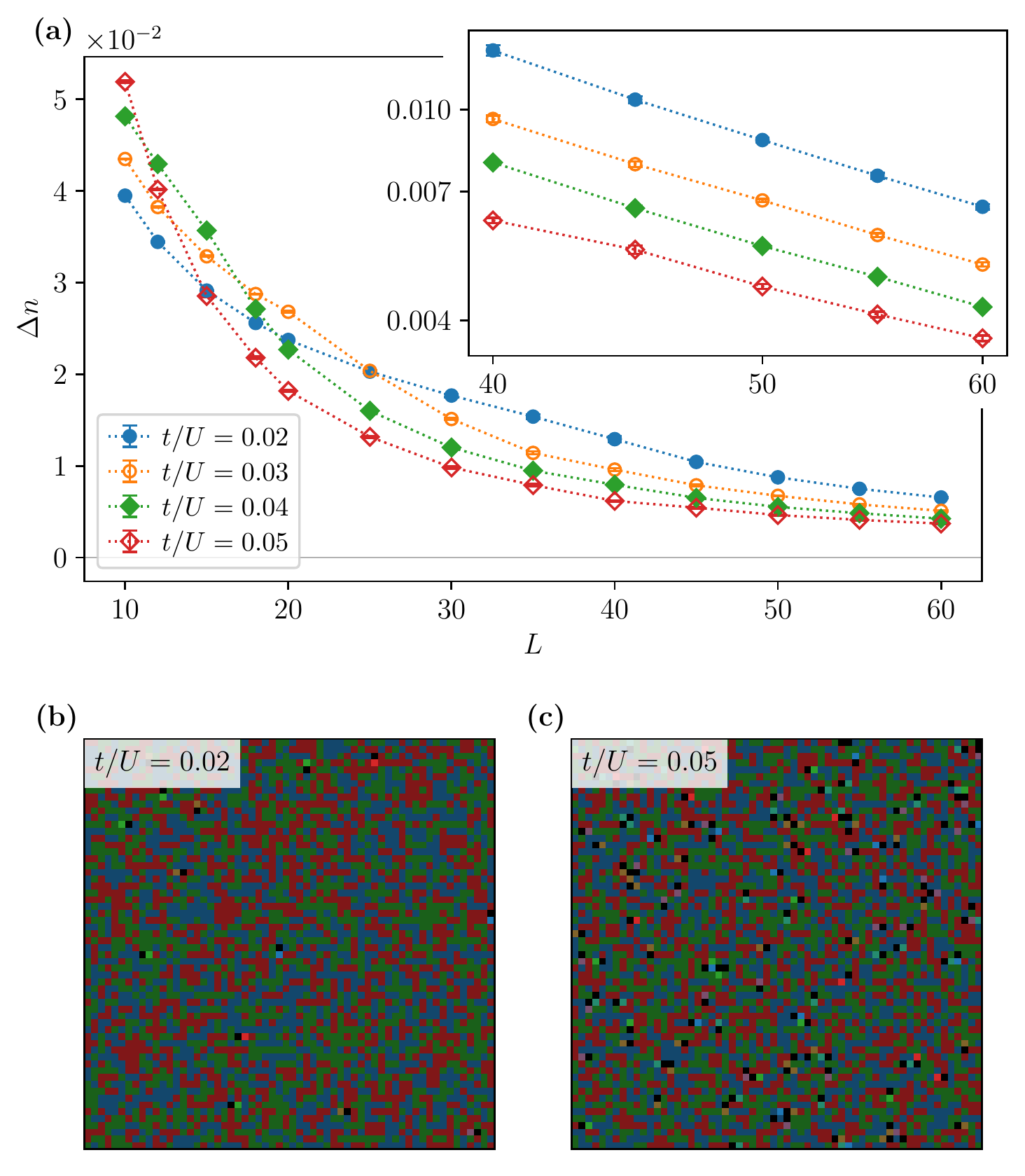}
    \caption{
        In panel {\bf (a)} the particle number-density imbalance $ \Delta n $, Eq.~(\ref{imbalance}), is plotted versus the system size $ L $ for various parameter values $ t/U $ with the inverse temperature given by $ \beta = L/t $.
        The total particle number-density $ n_\text{r} +  n_\text{g} + n_\text{b} = 1 $ is fixed while the particle number of each of the three symmetric components is allowed to fluctuate.
        The plot reveals a decay of $ \Delta n $ with increasing $ L $.
        The latter is more apparent in the inset which presents $ \Delta n $ for the largest $ L $s on logarithmic scales.
        %\sout{: the seemingly asymptotic linear trend implies an asymptotic power-law decay of $ \Delta n $ with $ 1/L $. } 
        This shows that $ \Delta n \rightarrow 0 $ as $ L \rightarrow \infty $.
        The bottom panels present two representative real-space particle-number distributions, i.e., imaginary-time slices of the world-line configuration, with $ L = 60 $ for $ t/U = 0.02 $ {\bf (b)} and $ t/U = 0.05 $ {\bf (c)}.
        The three different components are represented by the colors red, green, and blue, respectively. The darker the color, the fewer the particles occupying the corresponding site such that an empty site becomes black. These particle-numbers distributions clearly display a miscible phase which is corroborated by $ \Delta n $ of panel {\bf (a)}.
        In all cases the values $ U = 1 $ and $ U' = 0.9 $ were used.
    }
    \label{fig:particle_imbalance}
\end{figure}

Let us now characterize the absence of  phase separation in the system.
Namely, suppose the system spontaneously forms bound pairs between two components only, and one relaxes the restriction on individually fixed particle numbers $n_\alpha$. In that case, a disproportion between the components' particle densities is expected.
To that end, we investigate the density imbalance $ \Delta n $ defined through
\begin{equation}
    \langle \h n_\text{r} \h n_\text{g} \h n_\text{b} \rangle
    =
    (1/3 - 2 \Delta n)(1/3 + \Delta n)^2,
    \label{imbalance}
\end{equation}
while constraining the total particle-number density to $ n_\text{r} + n_\text{g} +  n_\text{b} = 1 $, and leaving $n_\alpha$ unrestricted such that only on average $ \langle \h n_\alpha \rangle = 1/3 $.
Here $ \h n_\alpha = \sum_i \h n_{i\alpha} / L^2 $.
If the particles of one component are completely exchanged in favor of particles belonging to the other two components, then $ \langle \h n_\text{r} \h n_\text{g} \h n_\text{b} \rangle = 0 $ leading to $ \Delta n = 1/6 $. Similarly, if two components are completely expelled then $ \Delta n = - 1/3 $. 
In contrast, if there is no density imbalance and all components are equally represented one should have $ \langle \h n_\text{r} \h n_\text{g} \h n_\text{b} \rangle = (1/3)^3 $ such that $ \Delta n = 0 $.
In Fig.~\ref{fig:particle_imbalance}~(a) we demonstrate a decay of $ \Delta n $ with increased system size $ L $ and inverse temperature $ \beta $ in the regime of the anticipated Borromean supercounterfluid phase.
As expected, we find a nonzero $ \Delta n $ at finite temperatures due to thermal density fluctuations, however, the magnitude of $ \Delta n $ is small in comparison to $ 1/3 $.
In addition, $ \Delta n $ show a clear decay with increased $ L $ indicating $ \Delta n \rightarrow 0 $ for $ L \rightarrow \infty $. This is further corroborated in Fig.~\ref{fig:particle_imbalance}~(b,c) by real-space particle-number distributions obtained from two representative imaginary-time slices of the world-line configurations.
This demonstrates that there is no density imbalance nor phase separation in the ground state.

\begin{figure}[t]
    \includegraphics[width=1\linewidth]{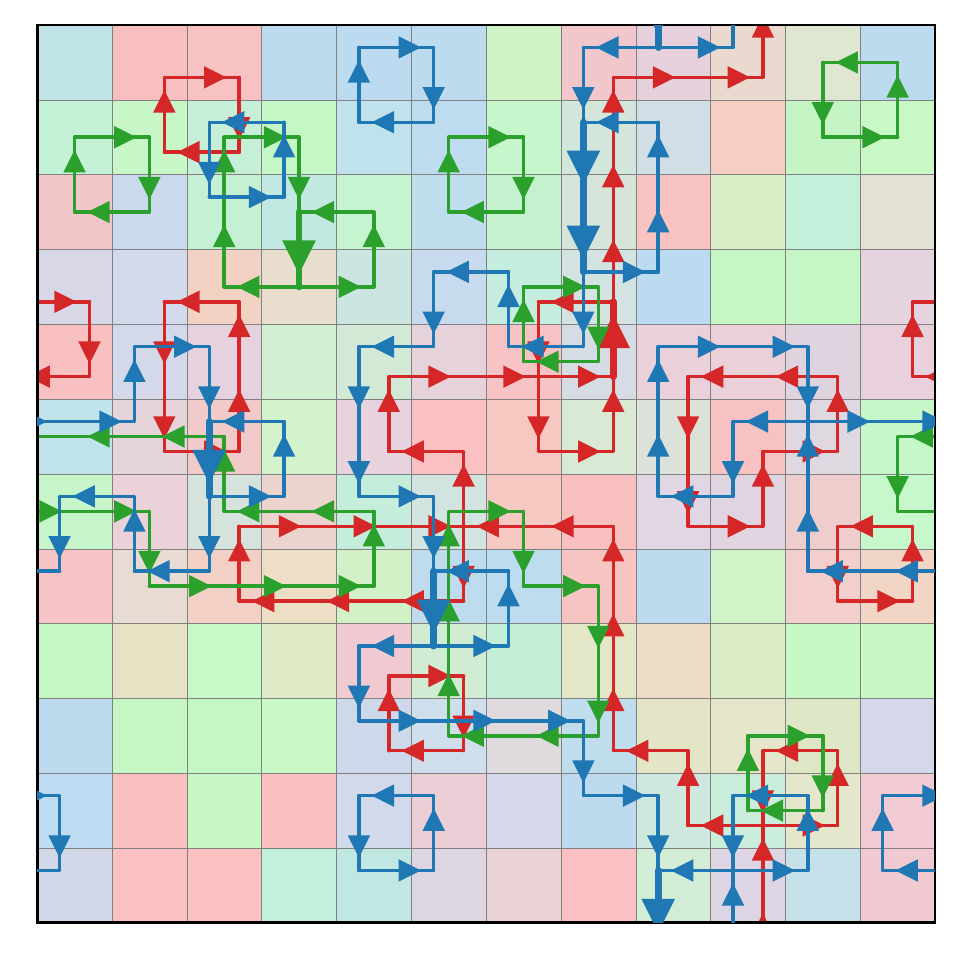}
    \caption{
    Spatial projection of a typical partition function world-line configuration of the Borromean supercounterfluid phase.
    Here the three different components are represented by red, green, and blue, respectively.
    The directional loops indicate the net particle propagation in imaginary time between adjacent sites, and the site's color indicates its average particle number population.
    The sites reveal three miscible components, and the loops display on average counter-propagating world-line trajectories. These properties are indicative of the Borromean supercounterfluid phase.
    By counting the flux of loops across the boundaries we find the winding numbers $ \mathbf w_\text{r} = \mathbf e_y $, $ \mathbf w_\text{g} = \mathbf 0 $, $ \mathbf w_\text{b} = -\mathbf e_y $ which satisfy $ \mathbf w_\text{r} + \mathbf w_\text{g} + \mathbf w_\text{b} = \mathbf 0 $.
    The results were obtained using $ t = 0.04 $, $ U = 1 $, $ U' = 0.9 $, $ L = 12 $, $ T = t/L $, and a fixed particle number density $ n_\alpha = 1/3 $.
    }
    \label{fig:loops}
\end{figure}

The Borromean supercounterfluid phase can be further studied by inspecting a typical partition function world-line configurations sampled with the help of the worm-algorithm Monte Carlo method. A representative configuration of the Borromean superfluid phase is illustrated in Fig.~\ref{fig:loops}, which indeed reveals three mixed components.
The net flow of particles in imaginary time further exhibits the  counterflowlike behavior, resulting in the winding numbers $ \langle \vec w_\alpha^2 \rangle \neq 0 $ and $ \langle (\sum_\alpha \vec w_\alpha )^2 \rangle = 0 $, or alternatively $ \rho_\text{d}/\rho_\text{k} = -1/2 $.

In conclusion, we have demonstrated microscopically that a strongly correlated three-component bosonic mixture---realizable in optical lattice setups---has a phase with  ``super" transport phenomenon different from conventional superfluidity.
In this phase, the simultaneous co-flow of all three bosonic components is arrested, while the system retains dissipationless counterflows between any pair of components.
These three counterflows are not independent but rather described by two superfluid degrees of freedom. At the microscopic level, the types of counterpropagating partners can vary.

Possible realization of these states in optical lattices could be obtained by trapping mixtures of bosonic isotopes of Na and K.
Possible ways to detect the Borromean supercounterfluid state experimentally is through tilting the lattice and detecting the ratio between transport of different individual components. However, the more striking signature can be obtained by observing a dramatic change in the system's rotational response.
Namely, a rotating conventional superfluid can be described by introducing a fictitious vector potential $ \boldsymbol{\Theta} $, leading to a vortex lattice formation (see, e.g., chapter 1 in \cite{svistunov2015superfluid}).
In the Borromean supercounterfluid state the vector potential $ \boldsymbol{\Theta} $ couples to the phase gradients through mass differences according to $f \approx \sum_{\alpha,\beta\neq\alpha}
[\nabla \theta_\alpha-\nabla \theta_\beta-(m_\alpha-m_\beta)\boldsymbol{\Theta}]^2$.
If all the components have equal particle masses, i.e., $m_\alpha = m_\beta$, the system is unaffected by rotation. Indeed, in that case, the counterflow involves no mass transfer, and vortices carry no angular momentum. On the other hand, if $ m_\alpha \neq m_\beta $ the system forms a vortex lattice to compensate for the superextensive increase of the free energy due to the rotation. In the latter case, the critical velocities and number of rotation-induced vortices will be proportional to the mass difference $ m_\alpha - m_\beta $ rather than to masses of individual components.

\section*{Acknowledgements}
We would like to thank Martin Zwierlein and Immanuel Bloch for discussions on experimental realizations of bosonic mixtures. E. Bl. 
E. Bl. and E. Ba. were supported by the Swedish Research Council Grants No. 2016-06122, 2018-03659, and G\"{o}ran Gustafsson Foundation for Research in Natural Sciences.
A. S. and E. Ba. acknowledge the support from Olle Engkvists stiftelse.
The computations were enabled by resources provided by the Swedish National Infrastructure for Computing (SNIC) at the National Supercomputer Centre (NSC) partially funded by the Swedish Research Council through grant agreement no. 2018-05973.

\bibliographystyle{apsrev4-1}
\bibliography{main}

\begin{thebibliography}{39}%
\makeatletter
\providecommand \@ifxundefined [1]{%
 \@ifx{#1\undefined}
}%
\providecommand \@ifnum [1]{%
 \ifnum #1\expandafter \@firstoftwo
 \else \expandafter \@secondoftwo
 \fi
}%
\providecommand \@ifx [1]{%
 \ifx #1\expandafter \@firstoftwo
 \else \expandafter \@secondoftwo
 \fi
}%
\providecommand \natexlab [1]{#1}%
\providecommand \enquote  [1]{``#1''}%
\providecommand \bibnamefont  [1]{#1}%
\providecommand \bibfnamefont [1]{#1}%
\providecommand \citenamefont [1]{#1}%
\providecommand \href@noop [0]{\@secondoftwo}%
\providecommand \href [0]{\begingroup \@sanitize@url \@href}%
\providecommand \@href[1]{\@@startlink{#1}\@@href}%
\providecommand \@@href[1]{\endgroup#1\@@endlink}%
\providecommand \@sanitize@url [0]{\catcode `\\12\catcode `\$12\catcode
  `\&12\catcode `\#12\catcode `\^12\catcode `\_12\catcode `\%12\relax}%
\providecommand \@@startlink[1]{}%
\providecommand \@@endlink[0]{}%
\providecommand \url  [0]{\begingroup\@sanitize@url \@url }%
\providecommand \@url [1]{\endgroup\@href {#1}{\urlprefix }}%
\providecommand \urlprefix  [0]{URL }%
\providecommand \Eprint [0]{\href }%
\providecommand \doibase [0]{http://dx.doi.org/}%
\providecommand \selectlanguage [0]{\@gobble}%
\providecommand \bibinfo  [0]{\@secondoftwo}%
\providecommand \bibfield  [0]{\@secondoftwo}%
\providecommand \translation [1]{[#1]}%
\providecommand \BibitemOpen [0]{}%
\providecommand \bibitemStop [0]{}%
\providecommand \bibitemNoStop [0]{.\EOS\space}%
\providecommand \EOS [0]{\spacefactor3000\relax}%
\providecommand \BibitemShut  [1]{\csname bibitem#1\endcsname}%
\let\auto@bib@innerbib\@empty
%</preamble>
\bibitem [{\citenamefont {Jaksch}\ \emph {et~al.}(1998)\citenamefont {Jaksch},
  \citenamefont {Bruder}, \citenamefont {Cirac}, \citenamefont {Gardiner},\
  and\ \citenamefont {Zoller}}]{jaksch1998cold}%
  \BibitemOpen
  \bibfield  {author} {\bibinfo {author} {\bibfnamefont {D.}~\bibnamefont
  {Jaksch}}, \bibinfo {author} {\bibfnamefont {C.}~\bibnamefont {Bruder}},
  \bibinfo {author} {\bibfnamefont {J.~I.}\ \bibnamefont {Cirac}}, \bibinfo
  {author} {\bibfnamefont {C.~W.}\ \bibnamefont {Gardiner}}, \ and\ \bibinfo
  {author} {\bibfnamefont {P.}~\bibnamefont {Zoller}},\ }\href@noop {}
  {\bibfield  {journal} {\bibinfo  {journal} {Physical Review Letters}\
  }\textbf {\bibinfo {volume} {81}},\ \bibinfo {pages} {3108} (\bibinfo {year}
  {1998})}\BibitemShut {NoStop}%
\bibitem [{\citenamefont {Jaksch}\ and\ \citenamefont
  {Zoller}(2005)}]{jaksch2005cold}%
  \BibitemOpen
  \bibfield  {author} {\bibinfo {author} {\bibfnamefont {D.}~\bibnamefont
  {Jaksch}}\ and\ \bibinfo {author} {\bibfnamefont {P.}~\bibnamefont
  {Zoller}},\ }\href {\doibase https://doi.org/10.1016/j.aop.2004.09.010}
  {\bibfield  {journal} {\bibinfo  {journal} {Annals of Physics}\ }\textbf
  {\bibinfo {volume} {315}},\ \bibinfo {pages} {52 } (\bibinfo {year}
  {2005})},\ \bibinfo {note} {special Issue}\BibitemShut {NoStop}%
\bibitem [{\citenamefont {Greiner}\ \emph {et~al.}(2002)\citenamefont
  {Greiner}, \citenamefont {Mandel}, \citenamefont {Esslinger}, \citenamefont
  {H{\"a}nsch},\ and\ \citenamefont {Bloch}}]{greiner2002quantum}%
  \BibitemOpen
  \bibfield  {author} {\bibinfo {author} {\bibfnamefont {M.}~\bibnamefont
  {Greiner}}, \bibinfo {author} {\bibfnamefont {O.}~\bibnamefont {Mandel}},
  \bibinfo {author} {\bibfnamefont {T.}~\bibnamefont {Esslinger}}, \bibinfo
  {author} {\bibfnamefont {T.~W.}\ \bibnamefont {H{\"a}nsch}}, \ and\ \bibinfo
  {author} {\bibfnamefont {I.}~\bibnamefont {Bloch}},\ }\href
  {http://dx.doi.org/10.1038/415039a} {\bibfield  {journal} {\bibinfo
  {journal} {Nature}\ }\textbf {\bibinfo {volume} {415}},\ \bibinfo {pages}
  {39} (\bibinfo {year} {2002})}\BibitemShut {NoStop}%
\bibitem [{\citenamefont {Bloch}(2005)}]{bloch2005ultracold}%
  \BibitemOpen
  \bibfield  {author} {\bibinfo {author} {\bibfnamefont {I.}~\bibnamefont
  {Bloch}},\ }\href@noop {} {\bibfield  {journal} {\bibinfo  {journal} {Nature
  physics}\ }\textbf {\bibinfo {volume} {1}},\ \bibinfo {pages} {23} (\bibinfo
  {year} {2005})}\BibitemShut {NoStop}%
\bibitem [{\citenamefont {Bloch}\ \emph {et~al.}(2008)\citenamefont {Bloch},
  \citenamefont {Dalibard},\ and\ \citenamefont {Zwerger}}]{bloch2008many}%
  \BibitemOpen
  \bibfield  {author} {\bibinfo {author} {\bibfnamefont {I.}~\bibnamefont
  {Bloch}}, \bibinfo {author} {\bibfnamefont {J.}~\bibnamefont {Dalibard}}, \
  and\ \bibinfo {author} {\bibfnamefont {W.}~\bibnamefont {Zwerger}},\
  }\href@noop {} {\bibfield  {journal} {\bibinfo  {journal} {Reviews of modern
  physics}\ }\textbf {\bibinfo {volume} {80}},\ \bibinfo {pages} {885}
  (\bibinfo {year} {2008})}\BibitemShut {NoStop}%
\bibitem [{\citenamefont {Lewenstein}\ \emph {et~al.}(2007)\citenamefont
  {Lewenstein}, \citenamefont {Sanpera}, \citenamefont {Ahufinger},
  \citenamefont {Damski}, \citenamefont {Sen},\ and\ \citenamefont
  {Sen}}]{lewenstein2007ultracold}%
  \BibitemOpen
  \bibfield  {author} {\bibinfo {author} {\bibfnamefont {M.}~\bibnamefont
  {Lewenstein}}, \bibinfo {author} {\bibfnamefont {A.}~\bibnamefont {Sanpera}},
  \bibinfo {author} {\bibfnamefont {V.}~\bibnamefont {Ahufinger}}, \bibinfo
  {author} {\bibfnamefont {B.}~\bibnamefont {Damski}}, \bibinfo {author}
  {\bibfnamefont {A.}~\bibnamefont {Sen}}, \ and\ \bibinfo {author}
  {\bibfnamefont {U.}~\bibnamefont {Sen}},\ }\href@noop {} {\bibfield
  {journal} {\bibinfo  {journal} {Advances in Physics}\ }\textbf {\bibinfo
  {volume} {56}},\ \bibinfo {pages} {243} (\bibinfo {year} {2007})}\BibitemShut
  {NoStop}%
\bibitem [{\citenamefont {Kuklov}\ and\ \citenamefont
  {Svistunov}(2003)}]{Kuklov2003counterflow}%
  \BibitemOpen
  \bibfield  {author} {\bibinfo {author} {\bibfnamefont {A.~B.}\ \bibnamefont
  {Kuklov}}\ and\ \bibinfo {author} {\bibfnamefont {B.~V.}\ \bibnamefont
  {Svistunov}},\ }\href {\doibase 10.1103/PhysRevLett.90.100401} {\bibfield
  {journal} {\bibinfo  {journal} {Phys. Rev. Lett.}\ }\textbf {\bibinfo
  {volume} {90}},\ \bibinfo {pages} {100401} (\bibinfo {year}
  {2003})}\BibitemShut {NoStop}%
\bibitem [{\citenamefont {Kuklov}\ \emph
  {et~al.}(2004{\natexlab{a}})\citenamefont {Kuklov}, \citenamefont
  {Prokof'ev},\ and\ \citenamefont {Svistunov}}]{kuklov2004commensurate}%
  \BibitemOpen
  \bibfield  {author} {\bibinfo {author} {\bibfnamefont {A.}~\bibnamefont
  {Kuklov}}, \bibinfo {author} {\bibfnamefont {N.}~\bibnamefont {Prokof'ev}}, \
  and\ \bibinfo {author} {\bibfnamefont {B.}~\bibnamefont {Svistunov}},\ }\href
  {\doibase 10.1103/PhysRevLett.92.050402} {\bibfield  {journal} {\bibinfo
  {journal} {Phys. Rev. Lett.}\ }\textbf {\bibinfo {volume} {92}},\ \bibinfo
  {pages} {050402} (\bibinfo {year} {2004}{\natexlab{a}})}\BibitemShut
  {NoStop}%
\bibitem [{\citenamefont {Kuklov}\ \emph
  {et~al.}(2004{\natexlab{b}})\citenamefont {Kuklov}, \citenamefont
  {Prokof'ev},\ and\ \citenamefont {Svistunov}}]{kuklov2004superfluid}%
  \BibitemOpen
  \bibfield  {author} {\bibinfo {author} {\bibfnamefont {A.}~\bibnamefont
  {Kuklov}}, \bibinfo {author} {\bibfnamefont {N.}~\bibnamefont {Prokof'ev}}, \
  and\ \bibinfo {author} {\bibfnamefont {B.}~\bibnamefont {Svistunov}},\ }\href
  {\doibase 10.1103/PhysRevLett.92.030403} {\bibfield  {journal} {\bibinfo
  {journal} {Phys. Rev. Lett.}\ }\textbf {\bibinfo {volume} {92}},\ \bibinfo
  {pages} {030403} (\bibinfo {year} {2004}{\natexlab{b}})}\BibitemShut
  {NoStop}%
\bibitem [{\citenamefont {Kuklov}\ \emph {et~al.}(2006)\citenamefont {Kuklov},
  \citenamefont {Prokof'ev}, \citenamefont {Svistunov},\ and\ \citenamefont
  {Troyer}}]{kuklov2006deconfined}%
  \BibitemOpen
  \bibfield  {author} {\bibinfo {author} {\bibfnamefont {A.}~\bibnamefont
  {Kuklov}}, \bibinfo {author} {\bibfnamefont {N.}~\bibnamefont {Prokof'ev}},
  \bibinfo {author} {\bibfnamefont {B.}~\bibnamefont {Svistunov}}, \ and\
  \bibinfo {author} {\bibfnamefont {M.}~\bibnamefont {Troyer}},\ }\href
  {http://dx.doi.org/10.1016/j.aop.2006.04.007} {\bibfield  {journal} {\bibinfo
   {journal} {Annals of Physics}\ }\textbf {\bibinfo {volume} {321}},\ \bibinfo
  {pages} {1602} (\bibinfo {year} {2006})}\BibitemShut {NoStop}%
\bibitem [{\citenamefont {Capogrosso-Sansone}\ \emph
  {et~al.}(2008)\citenamefont {Capogrosso-Sansone}, \citenamefont {S{\"o}yler},
  \citenamefont {Prokof'ev},\ and\ \citenamefont
  {Svistunov}}]{capogrosso2008monte}%
  \BibitemOpen
  \bibfield  {author} {\bibinfo {author} {\bibfnamefont {B.}~\bibnamefont
  {Capogrosso-Sansone}}, \bibinfo {author} {\bibfnamefont {{\c{S}}.~G.}\
  \bibnamefont {S{\"o}yler}}, \bibinfo {author} {\bibfnamefont
  {N.}~\bibnamefont {Prokof'ev}}, \ and\ \bibinfo {author} {\bibfnamefont
  {B.}~\bibnamefont {Svistunov}},\ }\href {\doibase 10.1103/PhysRevA.77.015602}
  {\bibfield  {journal} {\bibinfo  {journal} {Phys. Rev. A}\ }\textbf {\bibinfo
  {volume} {77}},\ \bibinfo {pages} {015602} (\bibinfo {year}
  {2008})}\BibitemShut {NoStop}%
\bibitem [{\citenamefont {Dahl}\ \emph
  {et~al.}(2008{\natexlab{a}})\citenamefont {Dahl}, \citenamefont {Babaev},
  \citenamefont {Kragset},\ and\ \citenamefont
  {Sudb\o{}}}]{dahl2008preemptive}%
  \BibitemOpen
  \bibfield  {author} {\bibinfo {author} {\bibfnamefont {E.~K.}\ \bibnamefont
  {Dahl}}, \bibinfo {author} {\bibfnamefont {E.}~\bibnamefont {Babaev}},
  \bibinfo {author} {\bibfnamefont {S.}~\bibnamefont {Kragset}}, \ and\
  \bibinfo {author} {\bibfnamefont {A.}~\bibnamefont {Sudb\o{}}},\ }\href
  {\doibase 10.1103/PhysRevB.77.144519} {\bibfield  {journal} {\bibinfo
  {journal} {Phys. Rev. B}\ }\textbf {\bibinfo {volume} {77}},\ \bibinfo
  {pages} {144519} (\bibinfo {year} {2008}{\natexlab{a}})}\BibitemShut
  {NoStop}%
\bibitem [{\citenamefont {Dahl}\ \emph
  {et~al.}(2008{\natexlab{b}})\citenamefont {Dahl}, \citenamefont {Babaev},\
  and\ \citenamefont {Sudb\o{}}}]{dahl2008unusual}%
  \BibitemOpen
  \bibfield  {author} {\bibinfo {author} {\bibfnamefont {E.~K.}\ \bibnamefont
  {Dahl}}, \bibinfo {author} {\bibfnamefont {E.}~\bibnamefont {Babaev}}, \ and\
  \bibinfo {author} {\bibfnamefont {A.}~\bibnamefont {Sudb\o{}}},\ }\href
  {\doibase 10.1103/PhysRevLett.101.255301} {\bibfield  {journal} {\bibinfo
  {journal} {Phys. Rev. Lett.}\ }\textbf {\bibinfo {volume} {101}},\ \bibinfo
  {pages} {255301} (\bibinfo {year} {2008}{\natexlab{b}})}\BibitemShut
  {NoStop}%
\bibitem [{\citenamefont {Dahl}\ \emph
  {et~al.}(2008{\natexlab{c}})\citenamefont {Dahl}, \citenamefont {Babaev},\
  and\ \citenamefont {Sudb\o{}}}]{Dahl2008hidden}%
  \BibitemOpen
  \bibfield  {author} {\bibinfo {author} {\bibfnamefont {E.~K.}\ \bibnamefont
  {Dahl}}, \bibinfo {author} {\bibfnamefont {E.}~\bibnamefont {Babaev}}, \ and\
  \bibinfo {author} {\bibfnamefont {A.}~\bibnamefont {Sudb\o{}}},\ }\href
  {\doibase 10.1103/PhysRevB.78.144510} {\bibfield  {journal} {\bibinfo
  {journal} {Phys. Rev. B}\ }\textbf {\bibinfo {volume} {78}},\ \bibinfo
  {pages} {144510} (\bibinfo {year} {2008}{\natexlab{c}})}\BibitemShut
  {NoStop}%
\bibitem [{\citenamefont {Capogrosso-Sansone}\ and\ \citenamefont
  {Kuklov}(2011)}]{capogrosso2011superfluidity}%
  \BibitemOpen
  \bibfield  {author} {\bibinfo {author} {\bibfnamefont {B.}~\bibnamefont
  {Capogrosso-Sansone}}\ and\ \bibinfo {author} {\bibfnamefont
  {A.}~\bibnamefont {Kuklov}},\ }\href@noop {} {\bibfield  {journal} {\bibinfo
  {journal} {Journal of Low Temperature Physics}\ }\textbf {\bibinfo {volume}
  {165}},\ \bibinfo {pages} {213} (\bibinfo {year} {2011})}\BibitemShut
  {NoStop}%
\bibitem [{\citenamefont {Sellin}\ and\ \citenamefont
  {Babaev}(2018)}]{sellin2018superfluid}%
  \BibitemOpen
  \bibfield  {author} {\bibinfo {author} {\bibfnamefont {K.}~\bibnamefont
  {Sellin}}\ and\ \bibinfo {author} {\bibfnamefont {E.}~\bibnamefont
  {Babaev}},\ }\href {\doibase 10.1103/PhysRevB.97.094517} {\bibfield
  {journal} {\bibinfo  {journal} {Phys. Rev. B}\ }\textbf {\bibinfo {volume}
  {97}},\ \bibinfo {pages} {094517} (\bibinfo {year} {2018})}\BibitemShut
  {NoStop}%
\bibitem [{\citenamefont {Babaev}()}]{babaev2002phase}%
  \BibitemOpen
  \bibfield  {author} {\bibinfo {author} {\bibfnamefont {E.}~\bibnamefont
  {Babaev}},\ }\href {https://arxiv.org/abs/cond-mat/0201547} {\bibinfo
  {journal} {arXiv preprint cond-mat/0201547}\ }\BibitemShut {NoStop}%
\bibitem [{\citenamefont {Babaev}\ \emph {et~al.}(2004)\citenamefont {Babaev},
  \citenamefont {Sudb{\o}},\ and\ \citenamefont
  {Ashcroft}}]{babaev2004superconductor}%
  \BibitemOpen
\bibfield  {journal} {  }\bibfield  {author} {\bibinfo {author} {\bibfnamefont
  {E.}~\bibnamefont {Babaev}}, \bibinfo {author} {\bibfnamefont
  {A.}~\bibnamefont {Sudb{\o}}}, \ and\ \bibinfo {author} {\bibfnamefont
  {N.}~\bibnamefont {Ashcroft}},\ }\href
  {http://dx.doi.org/10.1038/nature02910} {\bibfield  {journal} {\bibinfo
  {journal} {Nature}\ }\textbf {\bibinfo {volume} {431}},\ \bibinfo {pages}
  {666} (\bibinfo {year} {2004})}\BibitemShut {NoStop}%
\bibitem [{\citenamefont {Smiseth}\ \emph {et~al.}(2005)\citenamefont
  {Smiseth}, \citenamefont {Sm{\o}rgrav}, \citenamefont {Babaev},\ and\
  \citenamefont {Sudb{\o}}}]{smiseth2005field}%
  \BibitemOpen
  \bibfield  {author} {\bibinfo {author} {\bibfnamefont {J.}~\bibnamefont
  {Smiseth}}, \bibinfo {author} {\bibfnamefont {E.}~\bibnamefont
  {Sm{\o}rgrav}}, \bibinfo {author} {\bibfnamefont {E.}~\bibnamefont {Babaev}},
  \ and\ \bibinfo {author} {\bibfnamefont {A.}~\bibnamefont {Sudb{\o}}},\
  }\href {https://doi.org/10.1103/PhysRevB.71.214509} {\bibfield  {journal}
  {\bibinfo  {journal} {Physical Review B}\ }\textbf {\bibinfo {volume} {71}},\
  \bibinfo {pages} {214509} (\bibinfo {year} {2005})}\BibitemShut {NoStop}%
\bibitem [{\citenamefont {Herland}\ \emph {et~al.}(2010)\citenamefont
  {Herland}, \citenamefont {Babaev},\ and\ \citenamefont
  {Sudb{\o}}}]{herland2010phase}%
  \BibitemOpen
  \bibfield  {author} {\bibinfo {author} {\bibfnamefont {E.~V.}\ \bibnamefont
  {Herland}}, \bibinfo {author} {\bibfnamefont {E.}~\bibnamefont {Babaev}}, \
  and\ \bibinfo {author} {\bibfnamefont {A.}~\bibnamefont {Sudb{\o}}},\ }\href
  {https://doi.org/10.1103/PhysRevB.82.134511} {\bibfield  {journal} {\bibinfo
  {journal} {Physical Review B}\ }\textbf {\bibinfo {volume} {82}},\ \bibinfo
  {pages} {134511} (\bibinfo {year} {2010})}\BibitemShut {NoStop}%
\bibitem [{\citenamefont {Kuklov}\ \emph {et~al.}(2008)\citenamefont {Kuklov},
  \citenamefont {Matsumoto}, \citenamefont {Prokof'ev}, \citenamefont
  {Svistunov},\ and\ \citenamefont {Troyer}}]{kuklov2008deconfined}%
  \BibitemOpen
  \bibfield  {author} {\bibinfo {author} {\bibfnamefont {A.~B.}\ \bibnamefont
  {Kuklov}}, \bibinfo {author} {\bibfnamefont {M.}~\bibnamefont {Matsumoto}},
  \bibinfo {author} {\bibfnamefont {N.~V.}\ \bibnamefont {Prokof'ev}}, \bibinfo
  {author} {\bibfnamefont {B.~V.}\ \bibnamefont {Svistunov}}, \ and\ \bibinfo
  {author} {\bibfnamefont {M.}~\bibnamefont {Troyer}},\ }\href {\doibase
  10.1103/PhysRevLett.101.050405} {\bibfield  {journal} {\bibinfo  {journal}
  {Phys. Rev. Lett.}\ }\textbf {\bibinfo {volume} {101}},\ \bibinfo {pages}
  {050405} (\bibinfo {year} {2008})}\BibitemShut {NoStop}%
\bibitem [{\citenamefont {Agterberg}\ and\ \citenamefont
  {Tsunetsugu}(2008)}]{agterberg2008dislocations}%
  \BibitemOpen
  \bibfield  {author} {\bibinfo {author} {\bibfnamefont {D.}~\bibnamefont
  {Agterberg}}\ and\ \bibinfo {author} {\bibfnamefont {H.}~\bibnamefont
  {Tsunetsugu}},\ }\href {http://dx.doi.org/10.1038/nphys999} {\bibfield
  {journal} {\bibinfo  {journal} {Nature Physics}\ }\textbf {\bibinfo {volume}
  {4}},\ \bibinfo {pages} {639} (\bibinfo {year} {2008})}\BibitemShut {NoStop}%
\bibitem [{\citenamefont {Berg}\ \emph {et~al.}(2009)\citenamefont {Berg},
  \citenamefont {Fradkin},\ and\ \citenamefont {Kivelson}}]{berg2009charge}%
  \BibitemOpen
  \bibfield  {author} {\bibinfo {author} {\bibfnamefont {E.}~\bibnamefont
  {Berg}}, \bibinfo {author} {\bibfnamefont {E.}~\bibnamefont {Fradkin}}, \
  and\ \bibinfo {author} {\bibfnamefont {S.~A.}\ \bibnamefont {Kivelson}},\
  }\href {http://dx.doi.org/10.1038/nphys1389} {\bibfield  {journal} {\bibinfo
  {journal} {Nature Physics}\ }\textbf {\bibinfo {volume} {5}},\ \bibinfo
  {pages} {830} (\bibinfo {year} {2009})}\BibitemShut {NoStop}%
\bibitem [{\citenamefont {Weston}\ and\ \citenamefont
  {Babaev}(2021)}]{weston2021composite}%
  \BibitemOpen
  \bibfield  {author} {\bibinfo {author} {\bibfnamefont {D.}~\bibnamefont
  {Weston}}\ and\ \bibinfo {author} {\bibfnamefont {E.}~\bibnamefont
  {Babaev}},\ }\href {\doibase 10.1103/PhysRevB.104.075116} {\bibfield
  {journal} {\bibinfo  {journal} {Phys. Rev. B}\ }\textbf {\bibinfo {volume}
  {104}},\ \bibinfo {pages} {075116} (\bibinfo {year} {2021})}\BibitemShut
  {NoStop}%
\bibitem [{\citenamefont {Svistunov}\ \emph {et~al.}(2015)\citenamefont
  {Svistunov}, \citenamefont {Babaev},\ and\ \citenamefont
  {Prokof'ev}}]{svistunov2015superfluid}%
  \BibitemOpen
  \bibfield  {author} {\bibinfo {author} {\bibfnamefont {B.~V.}\ \bibnamefont
  {Svistunov}}, \bibinfo {author} {\bibfnamefont {E.~S.}\ \bibnamefont
  {Babaev}}, \ and\ \bibinfo {author} {\bibfnamefont {N.~V.}\ \bibnamefont
  {Prokof'ev}},\ }\href@noop {} {\emph {\bibinfo {title} {Superfluid states of
  matter}}}\ (\bibinfo  {publisher} {CRC Press},\ \bibinfo {address} {Boca
  Raton},\ \bibinfo {year} {2015})\BibitemShut {NoStop}%
\bibitem [{\citenamefont {Grinenko}\ \emph {et~al.}(2021)\citenamefont
  {Grinenko}, \citenamefont {Weston}, \citenamefont {Caglieris} \emph
  {et~al.}}]{grinenko2021bosonic}%
  \BibitemOpen
  \bibfield  {author} {\bibinfo {author} {\bibfnamefont {V.}~\bibnamefont
  {Grinenko}}, \bibinfo {author} {\bibfnamefont {D.}~\bibnamefont {Weston}},
  \bibinfo {author} {\bibfnamefont {F.}~\bibnamefont {Caglieris}},  \emph
  {et~al.},\ }\href {https://doi.org/10.1038/s41567-021-01350-9} {\bibfield
  {journal} {\bibinfo  {journal} {Nature Physics}\ } (\bibinfo {year}
  {2021})}\BibitemShut {NoStop}%
\bibitem [{\citenamefont {Andreev}\ and\ \citenamefont
  {Bashkin}(1975)}]{andreev1976three}%
  \BibitemOpen
  \bibfield  {author} {\bibinfo {author} {\bibfnamefont {A.~F.}\ \bibnamefont
  {Andreev}}\ and\ \bibinfo {author} {\bibfnamefont {E.~P.}\ \bibnamefont
  {Bashkin}},\ }\href {http://jetp.ras.ru/cgi-bin/e/index/e/42/1/p164?a=list}
  {\bibfield  {journal} {\bibinfo  {journal} {Soviet Physics JETP}\ }\textbf
  {\bibinfo {volume} {42}},\ \bibinfo {pages} {164} (\bibinfo {year}
  {1975})}\BibitemShut {NoStop}%
\bibitem [{\citenamefont {Linder}\ and\ \citenamefont
  {Sudb{\o}}(2009)}]{linder2009calculation}%
  \BibitemOpen
  \bibfield  {author} {\bibinfo {author} {\bibfnamefont {J.}~\bibnamefont
  {Linder}}\ and\ \bibinfo {author} {\bibfnamefont {A.}~\bibnamefont
  {Sudb{\o}}},\ }\href@noop {} {\bibfield  {journal} {\bibinfo  {journal}
  {Physical Review A}\ }\textbf {\bibinfo {volume} {79}},\ \bibinfo {pages}
  {063610} (\bibinfo {year} {2009})}\BibitemShut {NoStop}%
\bibitem [{\citenamefont {Hofer}\ \emph {et~al.}(2012)\citenamefont {Hofer},
  \citenamefont {Bruder},\ and\ \citenamefont {Stojanovi\ifmmode~\acute{c}\else
  \'{c}\fi{}}}]{hofer2012superfluid}%
  \BibitemOpen
  \bibfield  {author} {\bibinfo {author} {\bibfnamefont {P.~P.}\ \bibnamefont
  {Hofer}}, \bibinfo {author} {\bibfnamefont {C.}~\bibnamefont {Bruder}}, \
  and\ \bibinfo {author} {\bibfnamefont {V.~M.}\ \bibnamefont
  {Stojanovi\ifmmode~\acute{c}\else \'{c}\fi{}}},\ }\href {\doibase
  10.1103/PhysRevA.86.033627} {\bibfield  {journal} {\bibinfo  {journal} {Phys.
  Rev. A}\ }\textbf {\bibinfo {volume} {86}},\ \bibinfo {pages} {033627}
  (\bibinfo {year} {2012})}\BibitemShut {NoStop}%
\bibitem [{\citenamefont {Contessi}\ \emph {et~al.}(2021)\citenamefont
  {Contessi}, \citenamefont {Romito}, \citenamefont {Rizzi},\ and\
  \citenamefont {Recati}}]{contessi2021collisionless}%
  \BibitemOpen
  \bibfield  {author} {\bibinfo {author} {\bibfnamefont {D.}~\bibnamefont
  {Contessi}}, \bibinfo {author} {\bibfnamefont {D.}~\bibnamefont {Romito}},
  \bibinfo {author} {\bibfnamefont {M.}~\bibnamefont {Rizzi}}, \ and\ \bibinfo
  {author} {\bibfnamefont {A.}~\bibnamefont {Recati}},\ }\href@noop {}
  {\bibfield  {journal} {\bibinfo  {journal} {Physical Review Research}\
  }\textbf {\bibinfo {volume} {3}},\ \bibinfo {pages} {L022017} (\bibinfo
  {year} {2021})}\BibitemShut {NoStop}%
\bibitem [{\citenamefont {Hartman}\ \emph {et~al.}(2018)\citenamefont
  {Hartman}, \citenamefont {Erlandsen},\ and\ \citenamefont
  {Sudb\o{}}}]{Hartman2018}%
  \BibitemOpen
  \bibfield  {author} {\bibinfo {author} {\bibfnamefont {S.}~\bibnamefont
  {Hartman}}, \bibinfo {author} {\bibfnamefont {E.}~\bibnamefont {Erlandsen}},
  \ and\ \bibinfo {author} {\bibfnamefont {A.}~\bibnamefont {Sudb\o{}}},\
  }\href {\doibase 10.1103/PhysRevB.98.024512} {\bibfield  {journal} {\bibinfo
  {journal} {Phys. Rev. B}\ }\textbf {\bibinfo {volume} {98}},\ \bibinfo
  {pages} {024512} (\bibinfo {year} {2018})}\BibitemShut {NoStop}%
\bibitem [{\citenamefont {Syrwid}\ \emph {et~al.}(2021)\citenamefont {Syrwid},
  \citenamefont {Blomquist},\ and\ \citenamefont {Babaev}}]{vecdrag}%
  \BibitemOpen
  \bibfield  {author} {\bibinfo {author} {\bibfnamefont {A.}~\bibnamefont
  {Syrwid}}, \bibinfo {author} {\bibfnamefont {E.}~\bibnamefont {Blomquist}}, \
  and\ \bibinfo {author} {\bibfnamefont {E.}~\bibnamefont {Babaev}},\ }\href
  {\doibase 10.1103/PhysRevLett.127.100403} {\bibfield  {journal} {\bibinfo
  {journal} {Phys. Rev. Lett.}\ }\textbf {\bibinfo {volume} {127}},\ \bibinfo
  {pages} {100403} (\bibinfo {year} {2021})}\BibitemShut {NoStop}%
\bibitem [{\citenamefont {Gersch}\ and\ \citenamefont
  {Knollman}(1963)}]{gersch1963quantum}%
  \BibitemOpen
  \bibfield  {author} {\bibinfo {author} {\bibfnamefont {H.~A.}\ \bibnamefont
  {Gersch}}\ and\ \bibinfo {author} {\bibfnamefont {G.~C.}\ \bibnamefont
  {Knollman}},\ }\href {\doibase 10.1103/PhysRev.129.959} {\bibfield  {journal}
  {\bibinfo  {journal} {Phys. Rev.}\ }\textbf {\bibinfo {volume} {129}},\
  \bibinfo {pages} {959} (\bibinfo {year} {1963})}\BibitemShut {NoStop}%
\bibitem [{\citenamefont {Prokof'ev}\ \emph {et~al.}(1998)\citenamefont
  {Prokof'ev}, \citenamefont {Svistunov},\ and\ \citenamefont
  {Tupitsyn}}]{prokof1998worm}%
  \BibitemOpen
  \bibfield  {author} {\bibinfo {author} {\bibfnamefont {N.}~\bibnamefont
  {Prokof'ev}}, \bibinfo {author} {\bibfnamefont {B.}~\bibnamefont
  {Svistunov}}, \ and\ \bibinfo {author} {\bibfnamefont {I.}~\bibnamefont
  {Tupitsyn}},\ }\href {\doibase 10.1016/S0375-9601(97)00957-2} {\bibfield
  {journal} {\bibinfo  {journal} {Physics Letters A}\ }\textbf {\bibinfo
  {volume} {238}},\ \bibinfo {pages} {253 } (\bibinfo {year}
  {1998})}\BibitemShut {NoStop}%
\bibitem [{\citenamefont {Capogrosso-Sansone}\ \emph
  {et~al.}(2007)\citenamefont {Capogrosso-Sansone}, \citenamefont
  {Prokof’Ev},\ and\ \citenamefont {Svistunov}}]{capogrosso2007phase}%
  \BibitemOpen
  \bibfield  {author} {\bibinfo {author} {\bibfnamefont {B.}~\bibnamefont
  {Capogrosso-Sansone}}, \bibinfo {author} {\bibfnamefont {N.}~\bibnamefont
  {Prokof’Ev}}, \ and\ \bibinfo {author} {\bibfnamefont {B.}~\bibnamefont
  {Svistunov}},\ }\href@noop {} {\bibfield  {journal} {\bibinfo  {journal}
  {Physical Review B}\ }\textbf {\bibinfo {volume} {75}},\ \bibinfo {pages}
  {134302} (\bibinfo {year} {2007})}\BibitemShut {NoStop}%
\bibitem [{\citenamefont {Blomquist}(2021)}]{Blomquist1590738}%
  \BibitemOpen
  \bibfield  {author} {\bibinfo {author} {\bibfnamefont {E.}~\bibnamefont
  {Blomquist}},\ }\emph {\bibinfo {title} {Strong Correlation Effects in
  Bosonic and Fermionic Systems Through an Unbiased Quantum Monte Carlo
  Approach}},\ \href@noop {} {Ph.D. thesis},\ \bibinfo  {school} {KTH,
  Condensed Matter Theory} (\bibinfo {year} {2021})\BibitemShut {NoStop}%
\bibitem [{\citenamefont {Sellin}(2018)}]{SellinThesis}%
  \BibitemOpen
  \bibfield  {author} {\bibinfo {author} {\bibfnamefont {K.}~\bibnamefont
  {Sellin}},\ }\emph {\bibinfo {title} {Structure formation, phase transitions
  and drag interactions in multicomponent superconductors and superfluids}},\
  \href@noop {} {Ph.D. thesis},\ \bibinfo  {school} {KTH, Statistical Physics}
  (\bibinfo {year} {2018})\BibitemShut {NoStop}%
\bibitem [{\citenamefont {Lingua}\ \emph {et~al.}(2018)\citenamefont {Lingua},
  \citenamefont {Capogrosso-Sansone}, \citenamefont {Safavi-Naini},
  \citenamefont {Jahangiri},\ and\ \citenamefont
  {Penna}}]{lingua2018multiworm}%
  \BibitemOpen
  \bibfield  {author} {\bibinfo {author} {\bibfnamefont {F.}~\bibnamefont
  {Lingua}}, \bibinfo {author} {\bibfnamefont {B.}~\bibnamefont
  {Capogrosso-Sansone}}, \bibinfo {author} {\bibfnamefont {A.}~\bibnamefont
  {Safavi-Naini}}, \bibinfo {author} {\bibfnamefont {A.}~\bibnamefont
  {Jahangiri}}, \ and\ \bibinfo {author} {\bibfnamefont {V.}~\bibnamefont
  {Penna}},\ }\href@noop {} {\bibfield  {journal} {\bibinfo  {journal} {Physica
  Scripta}\ }\textbf {\bibinfo {volume} {93}},\ \bibinfo {pages} {105402}
  (\bibinfo {year} {2018})}\BibitemShut {NoStop}%
\bibitem [{\citenamefont {Pollock}\ and\ \citenamefont
  {Ceperley}(1987)}]{pollock1987path}%
  \BibitemOpen
  \bibfield  {author} {\bibinfo {author} {\bibfnamefont {E.~L.}\ \bibnamefont
  {Pollock}}\ and\ \bibinfo {author} {\bibfnamefont {D.~M.}\ \bibnamefont
  {Ceperley}},\ }\href {\doibase 10.1103/PhysRevB.36.8343} {\bibfield
  {journal} {\bibinfo  {journal} {Phys. Rev. B}\ }\textbf {\bibinfo {volume}
  {36}},\ \bibinfo {pages} {8343} (\bibinfo {year} {1987})}\BibitemShut
  {NoStop}%
\end{thebibliography}%

\end{document}